\documentstyle [preprint,aps]{revtex}
\input{epsf}

\def \biz   {\begin{itemize}}
\def \eiz   {\end{itemize}}

\begin{document}
\title{\bf{
Asymmetries Between the Production of ${D_s}^-$ and 
 ${D_s}^+$ Mesons from 500 GeV/$c$ ${\pi}^-$ Nucleon Interactions
as Functions of $x_F$ and ${p_t}^{2}$ }} 
\author{
    E.~M.~Aitala,$^9$
       S.~Amato,$^1$
    J.~C.~Anjos,$^1$
    J.~A.~Appel,$^5$
       D.~Ashery,$^{15}$
       S.~Banerjee,$^5$
       I.~Bediaga,$^1$
       G.~Blaylock,$^{8}$
    S.~B.~Bracker,$^{16}$
    P.~R.~Burchat,$^{14}$
    R.~A.~Burnstein,$^6$
       T.~Carter,$^5$
 H.~S.~Carvalho,$^{1}$
  N.~K.~Copty,$^{13}$
    L.~M.~Cremaldi,$^9$
 C.~Darling,$^{19}$
       K.~Denisenko,$^5$
       A.~Fernandez,$^{12}$
       P.~Gagnon,$^2$
       C.~Gobel,$^1$
       K.~Gounder,$^9$
     A.~M.~Halling,$^5$
       G.~Herrera,$^4$
 G.~Hurvits,$^{15}$
       C.~James,$^5$
    P.~A.~Kasper,$^6$
       S.~Kwan,$^5$
    D.~C.~Langs,$^{11}$
       J.~Leslie,$^2$
       B.~Lundberg,$^5$
       S.~MayTal-Beck,$^{15}$
       B.~Meadows,$^3$
 J.~R.~T.~de~Mello~Neto,$^1$
    D.~Mihalcea,$^7$
    R.~H.~Milburn,$^{17}$
 J.~M.~de~Miranda,$^1$
       A.~Napier,$^{17}$
       A.~Nguyen,$^7$
  A.~B.~d'Oliveira,$^{12}$
       K.~O'Shaughnessy,$^2$
    K.~C.~Peng,$^6$
    L.~P.~Perera,$^3$
    M.~V.~Purohit,$^{13}$
       B.~Quinn,$^8$
       S.~Radeztsky,$^{18}$
       A.~Rafatian,$^9$
    N.~W.~Reay,$^7$
    J.~J.~Reidy,$^9$
    A.~C.~dos Reis,$^1$
    H.~A.~Rubin,$^6$
 A.~K.~S.~Santha,$^3$
 A.~F.~S.~Santoro,$^1$
       A.~J.~Schwartz,$^{11}$
       M.~Sheaff,$^{18}$
    R.~A.~Sidwell,$^7$
    A.~J.~Slaughter,$^{19}$
    M.~D.~Sokoloff,$^3$
       N.~R.~Stanton,$^7$
       K.~Stenson,$^{18}$
    D.~J.~Summers,$^9$
 S.~Takach,$^{19}$
       K.~Thorne,$^5$
    A.~K.~Tripathi,$^{10}$
       S.~Watanabe,$^{18}$
 R.~Weiss-Babai,$^{15}$
       J.~Wiener,$^{11}$
       N.~Witchey,$^7$
       E.~Wolin,$^{19}$
       D.~Yi,$^9$
       S. Yoshida,$^{7}$                         
       R.~Zaliznyak,$^{14}$
       and
       C.~Zhang$^7$ \\
\begin{center} (Fermilab E791 Collaboration) \end{center}
}
\address{
$^1$ Centro Brasileiro de Pesquisas F{\'\i}sicas, Rio de Janeiro, Brazil\\
$^2$ University of California, Santa Cruz, California 95064\\
$^3$ University of Cincinnati, Cincinnati, Ohio 45221\\
$^4$ CINVESTAV, Mexico\\
$^5$ Fermilab, Batavia, Illinois 60510\\
$^6$ Illinois Institute of Technology, Chicago, Illinois 60616\\
$^7$ Kansas State University, Manhattan, Kansas 66506\\
$^{8}$ University of Massachusetts, Amherst, Massachusetts 01003\\
$^9$ University of Mississippi, University, Mississippi 38677\\
$^{10}$ The Ohio State University, Columbus, Ohio 43210\\
$^{11}$ Princeton University, Princeton, New Jersey 08544\\
$^{12}$ Universidad Autonoma de Puebla, Mexico\\
$^{13}$ University of South Carolina, Columbia, South Carolina 29208\\
$^{14}$ Stanford University, Stanford, California 94305\\
$^{15}$ Tel Aviv University, Tel Aviv, Israel\\
$^{16}$ 317 Belsize Drive, Toronto, Canada\\
$^{17}$ Tufts University, Medford, Massachusetts 02155\\
$^{18}$ University of Wisconsin, Madison, Wisconsin 53706\\
$^{19}$ Yale University, New Haven, Connecticut 06511\\
}

\date{\today}
\maketitle

\begin{abstract}

We present measurements of the production of ${D_s}^-$ mesons relative
to ${D_s}^+$ mesons as functions of $x_F$ and of ${p_t}^2$ for a sample
of 2445 $D_s$ decays to $\phi\pi$. The $D_s$ mesons were produced in
Fermilab experiment E791 with 500 GeV/$c$ $\pi^-$ mesons incident on one
platinum and four carbon foil targets. The acceptance-corrected
integrated asymmetry in the $x_F$ range $-0.1$ to 0.5 for ${D_s}^{\mp}$
mesons is 0.032 $\pm$ 0.022 $\pm$ 0.022, consistent with no net
asymmetry. We compare the results as functions of $x_F$ and ${p_t}^2$ to
predictions and to the large production asymmetry observed for $D^{\pm}$
mesons in the same experiment. These comparisons support the hypothesis
that production asymmetries come from the fragmentation process and not
from the charm quark production itself.
\end{abstract}
~\\
\clearpage
Previous studies of charm
particle production in hadron beams 
\cite{muon,na27,na32,wa82,e769,e769_ds,tom,WA89,beatrice}
have shown large enhancements in the
forward production of charm particles that contain a quark or di-quark
in common with the beam (leading particles) relative to those that do
not (non-leading particles). Neither leading-order
nor next-to-leading-order perturbative QCD calculations of charm quark
production can account for the observed asymmetries \cite{NDE,fnmr}.
The phenomenon is due either to unexpected contributions to charm quark
production or to features of the hadronization process. 

Three classes of models have been proposed to account for hadronization
in the beam fragmentation region: coalescence of produced charm quarks
with valence quarks or diquarks from the beam\cite{comb,bhk,mw},
coalescence of charm and valence quarks when both originate in the beam
particle \cite{brod,vogt,hwa}, and string fragmentation as implemented in the
Lund Model of the PYTHIA Monte Carlo program \cite{PYTHIA} and by
Piskounova \cite{Pisk}. All these models involve a mechanism of
attachment between the produced charm quarks and the valence quarks in
the beam, and include a varying amount of actual coalescence. This
coalescence is the dominant mechanism for the leading particle effect in
the first two classes of models. They include no mechanism which will
lead to a significant production asymmetry between particles except when
one of them has a quark in common with the beam and the other does not.
Since the probability of coalescence between two quarks is larger when
they have similiar velocities, the production asymmetry between leading
and nonleading charm mesons is expected to grow larger with increasing
$x_F$ and smaller with increasing ${p_t}^2$. While $D^\pm$ data from
Fermilab experiment E791 do show the expected increase in asymmetry 
with increasing $x_F$,
they do not support the prediction of decreasing asymmetry with
increasing ${p_t}^2$\cite{tom}.

In string fragmentation models, specifically the PYTHIA/LUND
implementation, a produced charm quark is always initially connected via
color strings to valence quarks or diquarks in the beam and the target.
A leading particle can be produced by a meson beam in cases where the
string attaches a charm quark to a beam anti-quark and/or an anticharm
quark to a beam quark. Some fraction of the time, the string invariant
mass is too small to allow the production of a quark-antiquark pair, and
the process reduces to that of final state coalescence, with the same
general features. However, in most cases additional $q\overline{q}$
pairs are produced in the evolution of the strings. The string
fragmentation model predicts an asymmetry even at relatively small
values of $x_F$. This asymmetry is predicted to be relatively flat as a
function of ${p_t}^2$, consistent with E791 measurements\cite{tom}.
String fragmentation also allows for a possible asymmetry in cases where
there is no quark in common with the beam, if one of the quarks is
produced in the string fragmentation. 

For the nearly pure ${\pi}^-(\overline{u}d)$ beam
\cite{anthony} in Fermilab experiment E791\cite{e791},
$D^-(\overline{c}d)$ is a leading particle and $D^+(c\overline{d})$ is
the corresponding nonleading particle. We have previously reported a
significant $D^{\mp}$ production asymmetry, increasing with $x_F$ and 
relatively flat in
${p_t}^2$, and well represented by a modified version of the PYTHIA/LUND Monte
Carlo\cite{tom}. In contrast, $D_{s}^+(c\overline{s})$ and
$D_{s}^-(\overline{c}s)$ mesons do not share any valence quark type with
the pion beam. There should thus be no asymmetry between ${D_s}^-$
and ${D_s}^+$ production if the effect is due to coalescence. The
limited $D_s$ sample sizes in previous experiments did not allow a
detailed measurement of the production asymmetry for $D_s$
mesons\cite{na32,e769Ds}. Indeed, no $D_{s}$ asymmetry data has been reported
for pion-induced production. We report the first results on the $D_s$
production asymmetry in pion-nucleon collisions, and with 
sufficient data to study
the asymmetry as functions of $x_{F}$ and of ${p_t}^{2}$.

The E791 spectrometer \cite{e791} had 23 silicon microstrip detector
planes (SMD's); six upstream of the target foils, seventeen downstream,
to measure charged particle trajectories. This SMD system allows precise
measurement of the position of the charm production vertex (primary
vertex) and charm decay vertex (secondary vertex). Two magnets, together
with the downstream SMD's, 35 drift chamber planes, and two proportional
wire chamber planes, provided momentum measurement. Particle
identification of pions, kaons, and protons was obtained with two
segmented, threshold \v{C}erenkov counters. The total E791 recorded data sample
is $2 \times 10^{10}$ 500 GeV/$c$ $\pi^-$ nucleon interactions, 
which satisfied a
loose requirement on the total transverse energy 
deposited in the calorimeter, assuming that particles originate at the
target foils.
After data reconstruction, only those interactions
in which one or more decay vertex candidates were detected were kept for
further analysis.

From the above sample, charm meson decays to $\phi\pi$ were sought by
starting with candidate $\phi\rightarrow K^+K^-$ decays. The $\phi$
selection began with combinations of two oppositely charged,
well-reconstructed tracks, at least one of which was identified as a
kaon by the \v{C}erenkov counters. The combination was retained as a
$\phi$ candidate if the invariant mass of the two tracks, assumed to be
kaons, was within 10 MeV/$c^2$ of the $\phi$ mass (1.0194 GeV/$c^2$).
For each $\phi$ candidate, a third track, assumed to be a pion, was
sought, for which the invariant mass of the three tracks, assumed to be
$KK\pi$, was within a mass window 1.79 - 2.05 GeV/$c^2$, which includes
both the $D_s$ and the $D^{\pm}$ masses. A fit to the point of
intersection of the three tracks was then performed.

To further distinguish charm meson decays from background, additional
selection criteria were applied to each secondary vertex candidate. The
values of the selection criteria were chosen to maximize the statistical
significance ($N_s/\sqrt{(N_s + N_b)}$) for each $x_F$ bin, where $N_s$
is the number of $D_s$ signal events and $N_b$ is the number of
background events under the $D_s$ mass peak. The signal is represented
by fully simulated ${D_s}$ events generated by the PYTHIA Monte Carlo
and the background is E791 data for which the candidate mass is in the
sideband of the $D_s$ peak, but outside the $D^+$ signal region. For
most of the $x_F$ range, typical cuts are as follows. The longitudinal
separation between the primary and secondary vertices was required to be
more than 10 times the experimental resolution for the measured
separation. The impact parameter of the reconstructed momentum vector of
the ${D_s}$ candidate with respect to the primary vertex was required to
be less than 40 microns. The ratio of the distances of closest approach
to the secondary and primary vertices were computed for each of the
three decay tracks. The product of the three ratios was required to be
less than 0.006. This means that in most cases the tracks that make the
secondary vertex should be significantly closer to the secondary vertex
than to the primary vertex. It was required that no track in the event,
except the three making the secondary vertex, pass within 10 microns of
the secondary vertex. 

Figure 1 shows the $K^+ K^- {\pi}^+$ and $K^+ K^- {\pi}^-$ invariant
mass distributions after the above selection criteria are applied. A fit
with two Gaussian curves and a linear background applied to the combined
spectrum yields 2445 $\pm$ 60 ${D_s}^{\pm}$ mesons and 1449 $\pm$ 55
$D^{\pm}$ mesons. Fits to the individual spectra yield 1188 $\pm$ 41
${D_s}^{+}$mesons, 1257 $\pm$ 42 ${D_s}^{-}$ mesons, 638 $\pm$ 37
${D}^{+}$ mesons and 811 $\pm$ 40 ${D}^{-}$ mesons. These uncorrected
data show a clear excess of ${D}^{-}$ mesons over ${D}^{+}$ mesons,
consistent with the earlier measurement by this collaboration for the
more copious decay $D^{\pm}\rightarrow K^{\mp}\pi^{\pm}\pi^{\pm}$
\cite{tom}, but show no statistically significant asymmetry in the
numbers of ${D_s}^-$ and ${D_s}^+$ mesons. 

In order to study the asymmetry in the production of ${D_s}^\pm$
mesons, $K^+K^-{\pi}^{\pm}$ invariant mass plots were made for
particular ranges of $x_F$ and ${p_t}^2$. The yields of ${D_s}^+$ and
${D_s}^-$ mesons were then measured by fitting the
peaks in the mass plots to Gaussian curves of fixed width and mean plus
a linear background, and then integrating the area under the Gaussian
curves. The evolution of the width of the Gaussian curves with $x_F$ and
${p_t}^2$ was determined from Monte Carlo simulation, and is the same
for both charge states.

 An asymmetry parameter $A$ for ${D_s}$ production is defined as
 
  $$  A \equiv \frac{ N_{{D_s}^-} - N_{{D_s}^+}}{N_{{D_s}^-} + N_{{D_s}^+}}$$ 
 Here, $N_{{D_s}^-}$  is the 
 number of negatively charged $D_s$ mesons and $N_{{D_s}^+}$ is the number of
 positively charged $D_s$ mesons 
 produced in a particular bin of $x_{F}$ or ${p_t}^{2}$. 

From studies using our complete Monte Carlo simulation, we find that our
acceptance, which includes the geometrical acceptance as well as the
reconstruction efficiency, falls off rapidly at high $x_F$. This
inefficiency at high $x_F$ increased with time during the data taking
period. This was mainly due to beam effects which caused a cumulative
local loss of efficiency in the drift chambers. A correction ranging
from $-6\%$ to +12\% was applied bin-by-bin for the difference between
the acceptances of positive and negative $D_s$ mesons. Residual
uncertainties in this correction are treated as systematic errors and
are taken from reference \cite{tom}, using the r.m.s. spread of four
Monte Carlo simulations representing different time periods during the
data taking. 

A second source of systematic uncertainty is due to $D_s$ mesons
produced by the 2.5\% $K^-(\overline{u}s)$ content of the incident beam
\cite{anthony},
for which the integrated asymmetry A is measured by E769, at 250 GeV/$c$
beam energy, to be $0.25\pm 0.11$\cite{e769_ds}. No correction to the
asymmetry is applied, but we assign a systematic error of 0.006 in the
$x_F$ range ($-0.1$ to 0.3) and 0.015 in the $x_F$ range (0.3-0.5) to
account for this source of systematic uncertainty. The increase in
systematic error at larger $x_F$ accounts for greater uncertainty
in the $K^-$ contribution in the forward direction.

The acceptance-corrected production asymmetry for $D_s$ mesons in 
the $x_F$ range
$-0.1$ to 0.5 is 0.032 $\pm$ 0.022 $\pm$ 0.022, consistent with no net
asymmetry. Here the first error is the statistical error and the second
error is the systematic error from the sources described above.

The acceptance-corrected asymmetries as functions of $x_F$ and ${p_t}^2$
are shown in Tables I and II and in Figures 2 and 3 respectively. A line
 at zero asymmetry is drawn in the two figures for reference. 
 Asymmetries in the production of ${D_s}^-$ and ${D_s}^+$ mesons as 
predicted by a tuned version of the PYTHIA/LUND Monte Carlo \cite{tuned}
are also shown in Figures 2 and 3 for comparison to the measured
asymmetries. In this version of PYTHIA, parameters such as the mass of
the charm quark and the intrinsic transverse momentum of the partons
have been tuned so that the asymmetry of the Monte Carlo sample matches
the asymmetry for the $D^+ \rightarrow K \pi \pi$ using E791 data
\cite{tom}. 
While E791 data are consistent with the PYTHIA model in both $x_F$ and 
${p_t}^2$,
a better $\chi^2$ is obtained for the hypothesis of $A$=0 in both cases. 
Figures 2 and 3
also show a comparison of $D_s$ and $D^{\pm}$ asymmetries as a function of
$x_F$ and ${p_t}^2$ measured using E791 data. 
The $D^{\pm}$ data are from a high
statistics measurement using the decay mode $D^+ \rightarrow K \pi \pi$
\cite{tom}. The production asymmetry is larger for $D^{\pm}$ mesons than
for $D_s$ mesons for all $x_F$ and ${p_t}^2$. An intrinsic
charm model predicts zero asymmetry for $D_s$ production at all
$x_F$ and ${p_t}^2$ \cite{vogt}. 

In summary, we have observed both $D^{\pm}$ and $D_s^{\pm}$ decays in
the same apparatus and find a net production asymmetry only for the
$D^{\pm}$. We find that the $D_s$ asymmetry is consistent with zero in
each bin of $x_F$ and ${p_t}^2$ in the ranges covered by E791. Since the
beam used in experiment E791 contains almost no strange particles and
the $D_s$ asymmetry is consistent with zero, it can be concluded that
the asymmetry seen in the $D^\pm$ mesons is not likely to come
from any asymmetry in the production of the charm quarks themselves, but
originates in the hadronization process.

        We gratefully acknowledge the assistance of the staffs of
Fermilab and of all the participating institutions. This research was
supported by the Brazilian Conselho Nacional de Desenvolvimento
Cient\'\i fico e Tecnol\'ogico, CONACyT (Mexico), the Israeli Academy of
Sciences and Humanities, the U.S. Department of Energy, the
U.S.-Israel Binational Science Foundation and the U.S. National
Science Foundation. Fermilab is operated by the Universities Research
Association, Inc., under contract with the United States Department of
Energy.

\bibliographystyle{unsrt}

\newpage
\noindent TABLE I. $D_s$ production asymmetry as a function of $x_F$ 
from E791 data,
integrated over the ${p_t}^{2}$ range 0 to 10
$(GeV/c)^2$. The first error is statistical and the second systematic.
\begin{center}
\begin{tabular}{c|c} \hline \hline
{$x_F$} & \hspace{0.35in} {$D_s$ asymmetry A} \\  \hline
$-0.1$ $-$ 0.0   & \hspace {0.35in} 0.015 $\pm$ 0.054 $\pm$ 0.023 \\ 
 0.0 $-$ 0.05 & \hspace{0.35in} $-0.002 \pm 0.045 \pm 0.007$ \\ 
 0.05 $-$ 0.1 & \hspace{0.35in} 0.001 $\pm$ 0.047 $\pm$ 0.020 \\ 
 0.1 $-$ 0.2  & \hspace{0.35in} 0.020 $\pm$ 0.042 $\pm$ 0.027 \\ 
 0.2 $-$ 0.25 & \hspace{0.35in} 0.080 $\pm$ 0.091 $\pm$ 0.018 \\ 
 0.25 $-$ 0.3 & \hspace{0.35in} 0.083 $\pm$ 0.125 $\pm$ 0.036 \\ 
 0.3 $-$ 0.4  & \hspace{0.35in} 0.104 $\pm$ 0.140 $\pm$ 0.075  \\ 
 0.4 $-$ 0.5  & \hspace{0.35in} 0.149 $\pm$ 0.252 $\pm$ 0.105  \\ \hline \hline
\end{tabular}
\end{center}
~\\
~\\
\noindent TABLE II. $D_s$ production asymmetry as a function 
of ${p_t}^{2}$ from E791 data,
integrated over the $x_F$ range $-0.1$ to 0.5. The first error is statistical
 and the second systematic. 
\begin{center}
\begin{tabular}{c|c} \hline \hline
  {${p_t}^2$(GeV/$c)^2$} & \hspace{0.35in}{$D_s$ asymmetry A}  \\  \hline
 0.0 $-$ 1.0 & \hspace{0.35in} $-0.016 \pm 0.035 \pm 0.013$\\ 
 1.0 $-$ 2.0 & \hspace{0.35in}  0.036 $\pm$ 0.043 $\pm$ 0.036 \\ 
 2.0 $-$ 3.0 & \hspace{0.35in}  0.089 $\pm$ 0.052 $\pm$ 0.056 \\ 
 3.0 $-$ 4.0 & \hspace{0.35in}  0.081 $\pm$ 0.056 $\pm$ 0.089 \\ 
 4.0 $-$ 7.0 & \hspace{0.35in}  0.032 $\pm$ 0.053 $\pm$ 0.078 \\ 
 7.0 $-$ 10.0 &\hspace{0.35in}$-0.077 \pm 0.053 \pm 0.068$ \\ \hline \hline
\end{tabular}
\end{center}
\pagebreak
\begin{figure}[p]
\centerline {\epsfxsize=6in  \epsffile{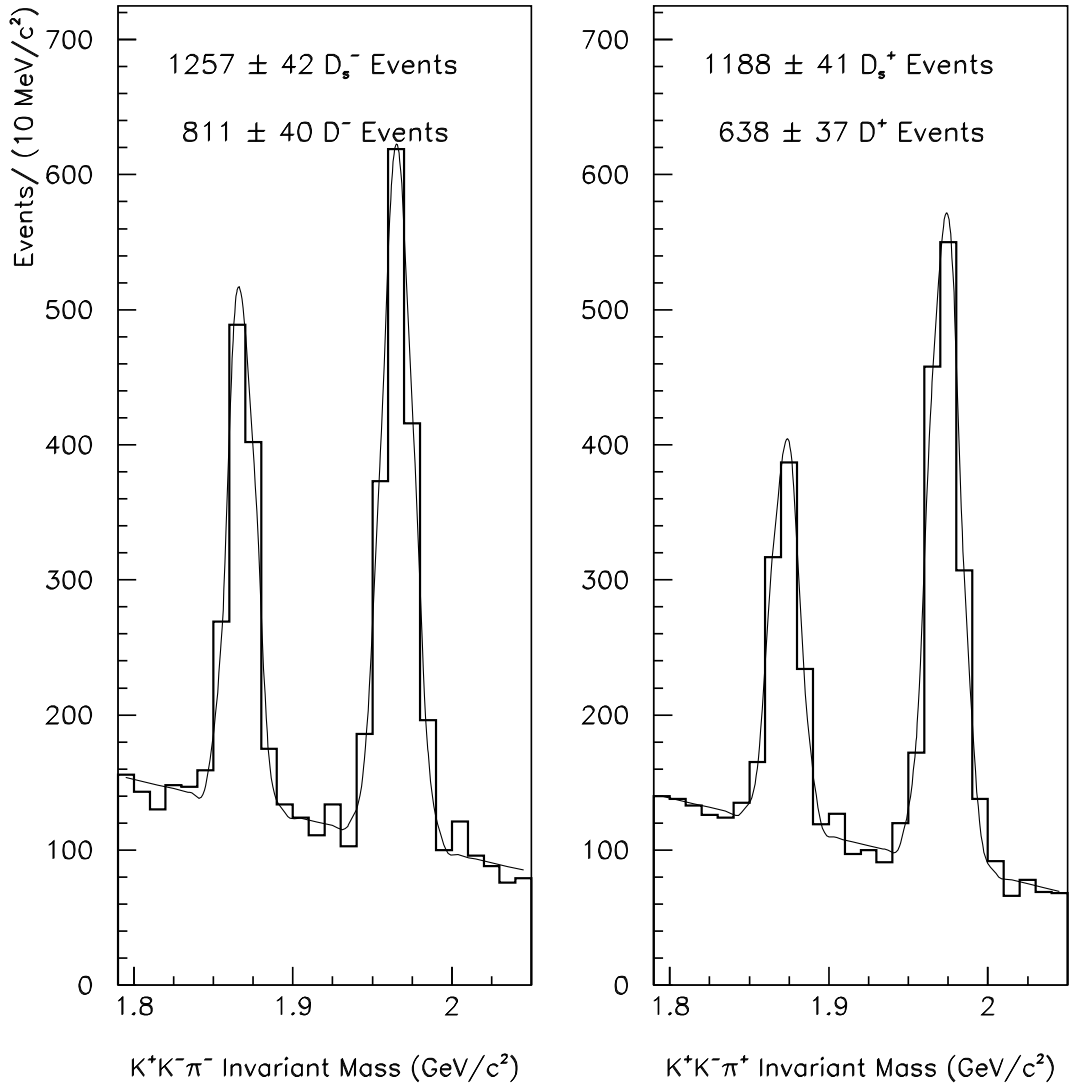} }
\caption{$K^+K^-\pi^-$ and $K^+K^-\pi^+$ invariant mass plots.}
\end{figure}
\newpage
\begin{figure}[p]
\centerline {\epsfxsize=6in  \epsffile{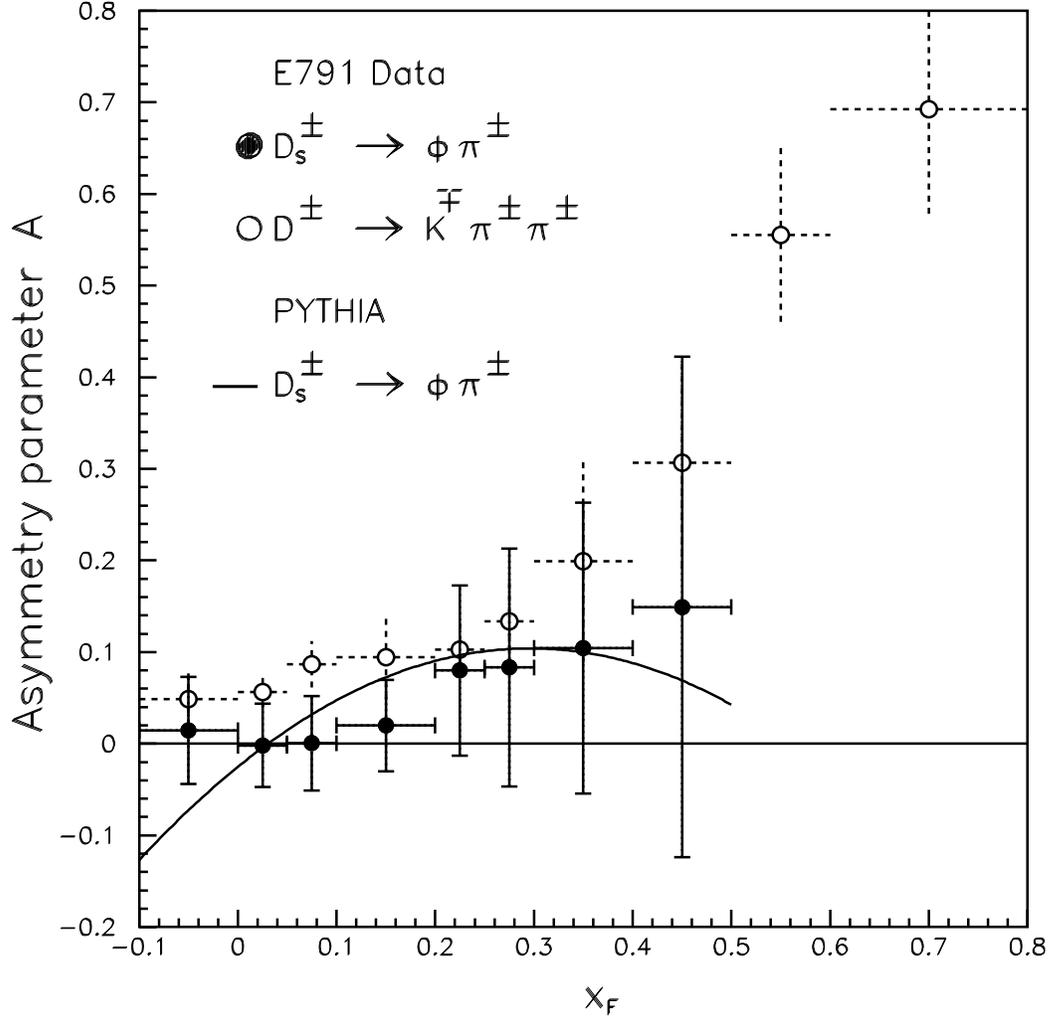} }
\caption{Comparison of the $D_s$ (solid circles) production asymmetry 
as a function of $x_F$ to the $D^+$ (open circles) production asymmetry, both
measured by experiment E791.
The acceptance-corrected data are integrated over the ${p_t}^2$ interval 
(1-10) (GeV/c)$^2$.
Also shown in the figure is the $D_s$ production asymmetry (solid line)
predicted by the ``tuned'' PYTHIA Monte Carlo described in the text.}
\end{figure}
\newpage
\begin{figure}[p]
\centerline {\epsfxsize=6in  \epsffile{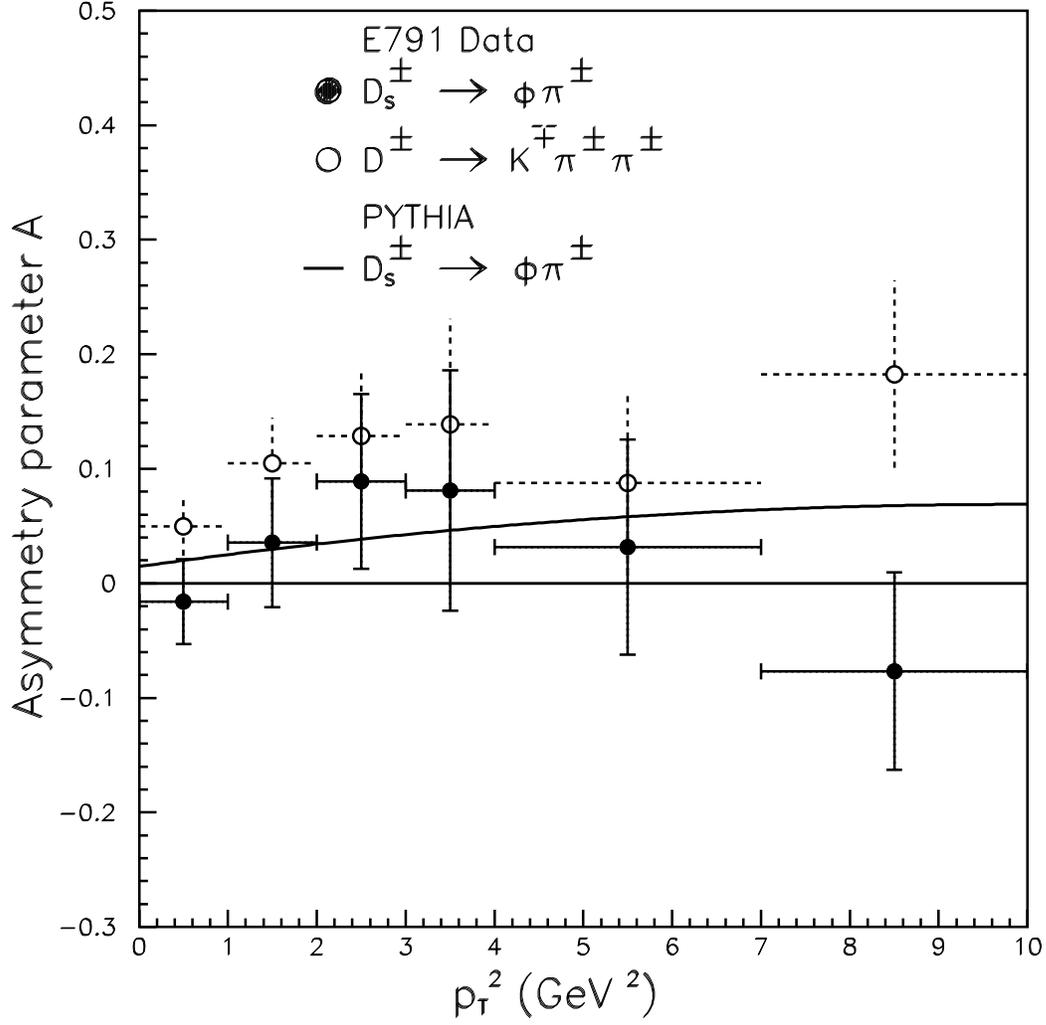} }
\caption{Comparison of the $D_s$ (solid circles) production asymmetry 
as a function of ${p_t}^2$ to the $D^+$ (open circles) production asymmetry,
both measured by experiment E791.
 The acceptance-corrected data are integrated over the $x_F$ interval
 ($-0.1$, 0.5) for $D_s$ mesons and ($-0.1$, 0.8) for $D^\pm$ mesons.
Also shown in the figure is the $D_s$ production asymmetry (solid line) 
predicted by the ``tuned'' PYTHIA Monte Carlo described in the text. }
\end{figure}
\end{document}